# Performance of TCP over ABR with Long-Range Dependent VBR Background Traffic over Terrestrial and Satellite ATM networks [1]

Shivkumar Kalyanaraman[3], Bobby Vandalore, Raj Jain, Rohit Goyal, Sonia Fahmy
The Ohio State University, Department of CIS
Columbus, OH 43210-1277
Phone: 614-292-3989, Fax: 614-292-2911 Email: {*shivkuma, vandalor, jain*}@cis.ohio-state.edu
Sastri Kota
Lockheed Martin Telecommunications,
1272 Borregas Avenue, Sunnyvale, CA 94088
Email: kota@lmsc.lockheed.com

**Abstract**

Compressed video is well known to be self-similar in nature [1, 2]. We model VBR carrying Long-Range Dependent (LRD), multiplexed MPEG-2 video sources. The actual traffic for the model is generated using fast-fourier transform of generate the fractional gaussian noise (FGN) sequence [14]. Our model of compressed video sources bears similarity to an MPEG-2 Transport Stream carrying video, i.e., it is long-range dependent [2] and generates traffic in a piecewise-CBR fashion [3]. We study the effect of such VBR traffic on ABR carrying TCP traffic. The effect of such VBR traffic is that the ABR capacity is highly variant. We find that a switch algorithm like ERICA+ [4] can tolerate this variance in ABR capacity while maintaining high throughput and low delay. We present simulation results for terrestrial and satellite configurations.

# 1 Introduction

The ABR model has been extensively studied with different source traffic patterns like persistent sources, ON-OFF bursty sources, ping pong sources, TCP sources, long-range dependent (or self-similar) sources, and source-bottlenecked VCs. Many of these studies have also considered the performance in the presence of ON-OFF VBR background traffic.

In reality, VBR consists of multiplexed compressed audio and video application traffic, each shaped by leaky buckets at their respective Sustained Cell Rate (SCR) and Peak Cell Rate (PCR) parameters. Compressed video has been shown to be long-range dependent by nature [1, 2]. Compressed audio and video streams belonging to a single program are expected to be carried over an ATM network using the MPEG-2 Transport Stream facility as outlined in reference [3].

In this paper, we first present a model of multiplexed MPEG-2 transport streams carried over ATM using the VBR service. Each stream exhibits long-range dependence, i.e., correlation over large time scales. We then study the effect of this VBR background on ABR connections carrying TCP file transfer applications on WAN and satellite configurations.

[3]Shivkumar Kalyanaraman is now with Dept. of ECSE, Rensselaer Polytechnic Institute, Troy, NY 12180-3590

## 2 Overview of MPEG-2 over ATM

In this section, we give a quick introduction to the MPEG-2 over ATM model and introduce some MPEG-2 terminology. For a detailed discussion, see reference [5].

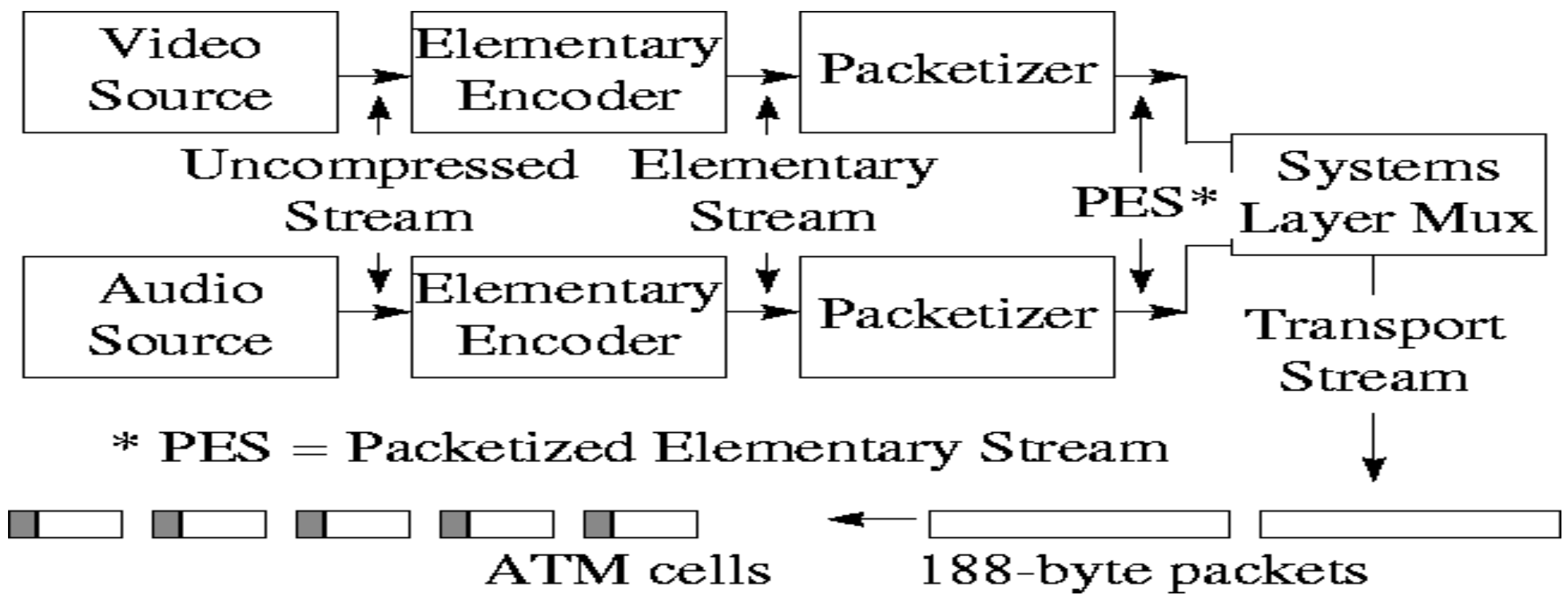

Figure 1: Overview of MPEG-2 Transport Streams

The MPEG-2 standard specifies two kinds of streams to carry coded video: the "Transport Stream" and the "Program Stream". The latter is used for compatibility with MPEG-1 (used for stored compressed video/audio), while the former is used to carry compressed video over networks which may not provide an end-to-end constant delay and jitter-free abstraction.

A Transport Stream can carry several programs multiplexed into one stream. Each program may consist of several "elementary streams," each containing MPEG-2 compressed video, audio, and other streams like close-captioned text, etc.

Figure 1 shows one such program stream formed by multiplexing a compressed video and a compressed audio elementary stream. Specifically, the figure shows the uncompressed video/audio stream going through the MPEG-2 elementary encoder to form the elementary stream. Typically, the uncompressed stream consists of frames generated at constant intervals (called "frame display times") of 33 ms (NTSC format) or 40 ms (PAL format). These frames (or "Group of Pictures" in MPEG-2 terminology) are called "Presentation Units." MPEG-2 compression produces three different types of frames: I, P and B frames, called "Access Units," as illustrated in Figure 2.

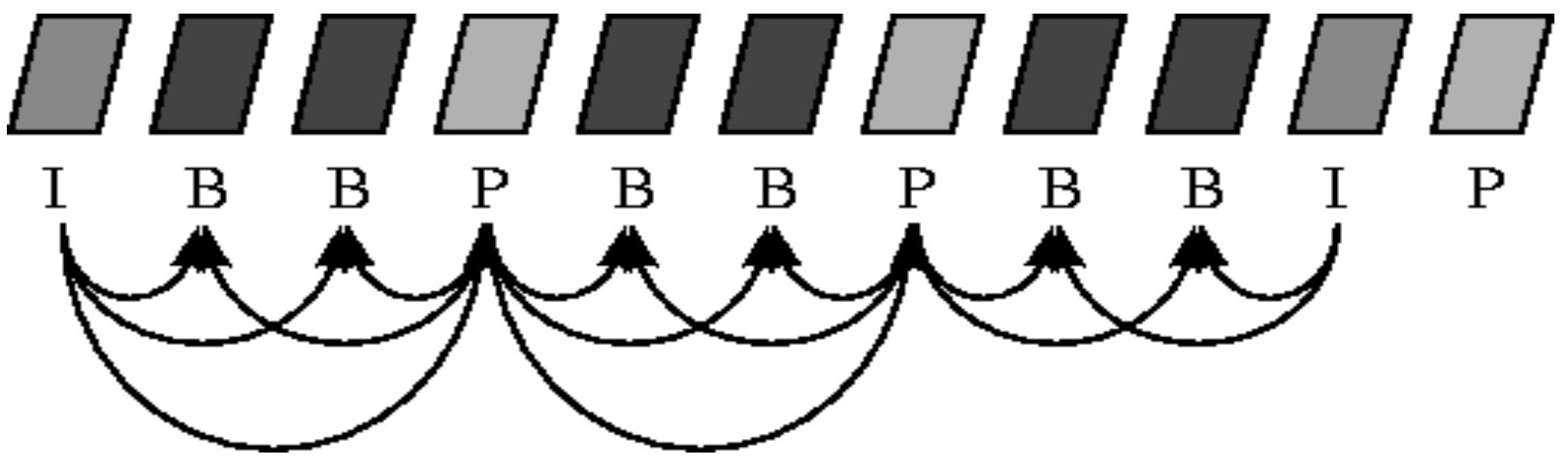

Figure 2: The I, P and B frames of MPEG-2

I (Intra-) frames are large. They contain the base picture, autonomously coded without need of a reference to another picture. They might take about 4-5 frame display times (approximately 160 ms) to be transmitted on the network depending upon the available rate [6].

P (Predictive-) frames are medium-sized. They are coded with respect to previous I or P frame. Transmission times for P frames is typically about 0.5-1 frame display times [6].

B (Birectionally predicted-) frames are very small. They are coded with respect to previous and later I or P frames and achieve maximum compression ratios (200:1). Transmission times for B frames is typically about 0.2 frame display times or even less [6].

As shown in Figure 1, the access units are packetized to form the "Packetized Elementary Stream (PES)". PES packets may be variable in length. The packetization process is implementation specific. PES packets may carry timestamps (called Presentation Timestamps (PTS) and Decoding Timestamps (DTS)) for long-term synchronization. The MPEG-2 standard specifies that PTS timestamps must appear at least once every 700 ms.

The next stage is the MPEG-2 Systems Layer which does the following four functions. First, it creates fixed size (188 byte) transport packets from PES packets. Second, the transport packets of different PESs belonging to one program are identified as such in the transport packet format. Third, it multiplexes several such programs to create a single Transport Stream. Fourth, it samples a system clock (running at 27 MHz) and encodes timestamps called "MPEG2 Program Clock References" (MPCRs, see [3]) in every multiplexed program. The time base for different programs may be different.

The MPCRs are used by the destination decoder to construct a Phase Locked Loop (PLL) and synchronize with the clock in the incoming stream. The MPEG-2 standard specifies that MPCRs must be generated at least once every 100 ms. Due to AAL5 packetization considerations, vendors usually also fix a maximum rate of generation of MPCRs to 50 per second (i.e. no less than one MPCR per 20 ms).

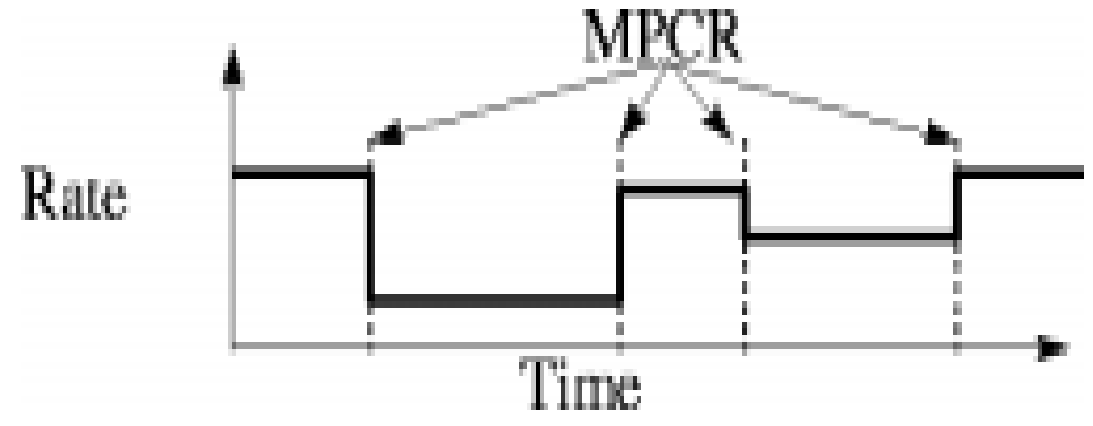

Figure 3: Piecewise constant nature of MPEG-2 Single Program Transport Streams

The key point is that the MPEG-2 rate is piecewise-CBR. As shown in Figure 3, the program's rate (not the transport stream's rate) is constant between successive MPCRs. The maximum rate is bounded by a peak value (typically 15 Mbps for HDTV quality compressed video [5]). The choice of the rates between MPCRs is implementation specific, but in general depends upon the buffer occupancy, and the rate of generation of the elementary streams.

The transport stream packets are encapsulated in AAL5 PDU with two transport stream packets in a single AAL5 PDU (for efficiency). The encapsulation method does not look for MPCRs in a transport packet and might introduce some jitter in the process. Alternate methods and enhancements to the above method have been proposed [5, 7].

An ATM VBR connection can multiplex several transport streams, each containing several programs, which in turn can contain several elementary streams. We model the multiplexing of several transport streams over VBR. But in our model, we will have only one program per Transport Stream (called the "Single Program Transport Stream" or "SPTS").

MPEG-2 uses a constant end-to-end delay model. The decoder at the destination can use techniques like having a de-jittering buffer, or restamping the MPCRs to compensate for network jitter, [5]. There is a Phase Locked Loop (PLL) at the destination which locks onto the MPCR clock in the incoming stream. The piecewise-CBR requirement allows the recovered clock to be reliable. Engineering of ATM VBR VCs to provide best service for MPEG-2 transport streams and negotiation of rates (PCR, SCR) is currently an important open question.

# 3 VBR Video modeling

There have been several attempts to model compressed video, see references [2, 8, 9] and references therein. Beran et al [2] show that long-range dependence is an inherent characteristic of compressed VBR video. But, they do not consider MPEG-2 data. Garrett and Willinger [8] show that a combination of distributions is needed to model VBR video. Heyman and Lakshman [9] argue that simple markov chain models are sufficient for traffic engineering purposes even though the frame size distribution may exhibit long-range dependence.

The video traffic on the network may be affected further by the multiplexing, renegotiation schemes, feedback schemes and the service category used. Examples of renegotiation, feedback schemes and best-effort video delivery are found in the literature, [10, 11, 12].

We believe that a general model of video traffic on the ATM network is yet to be discovered. In this paper, we are interested in the performance of ABR carrying TCP connections when affected by a long-range dependent, highly variable VBR background. We hence need a model for the video background. We have attempted to design the model to resemble the MPEG-2 Transport Stream.

There are three parameters in the model: the compressed video frame size, the inter-MPCR interval lengths, and the rates in these inter-MPCR intervals. In our model, the inter-MPCR intervals are uniformly distributed and the rates in the inter-MPCR intervals are long-range dependent. In real products, the rates are chosen depending upon the buffer occupancy at the encoder, which in turn depends upon the frame sizes of the latest set of frames generated. Further, the range of inter-MPCR intervals we generate follows implementation standards. We believe that this models the MPEG-2 Transport Stream, and still incorporates the long-range dependence property in the video streams. The effect of this VBR model on ABR is to introduce high variance in ABR capacity. As we shall see, the ERICA+ algorithm deals with the variance in ABR capacity and successfully bounds the maximum ABR queues, while maintaining high link utilization.

# 4 Modeling MPEG-2 Transport Streams over VBR

We model a "video source" as consisting of a transport stream generator, also called encoder (E) and a network element (NE). The encoder produces a Transport Stream as shown in Figure 1 and discussed in section 2. In our model, the Transport Stream consists of a single program stream. The network element encapsulates the transport packets into AAL5 PDUs and then fragments them into cells. The output of the network element (NE) goes to a leaky bucket which restricts the peak rate to 15 Mbps. This leaky bucket function can alternatively be done in the encoder, E (which does not send transport packets beyond a peak rate).

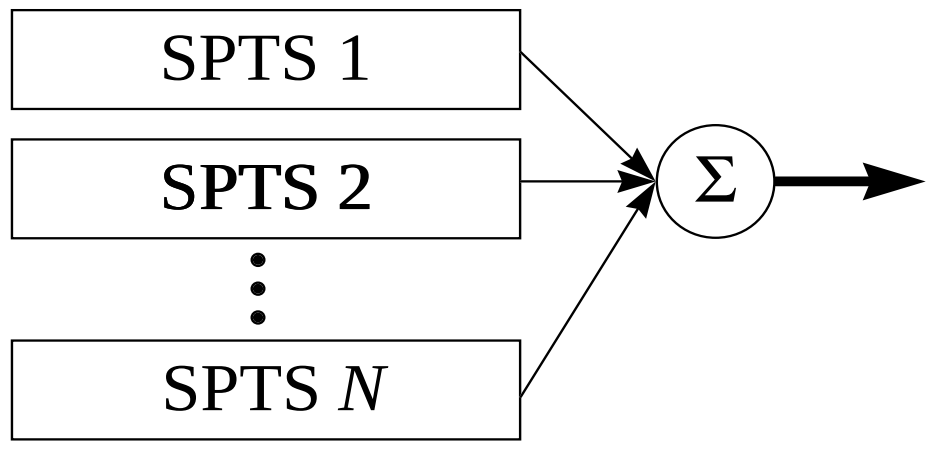

Figure 4: Multiplexing MPEG-2 Single Program Transport Streams (SPTSs) over VBR

Several (N) such video sources are multiplexed to form the VBR traffic going into the network as shown in Figure 4. Each encoder generates MPCRs uniformly distributed between 20 ms and 100 ms. The reason for this choice (of maximum and minimum MPCRs) is explained in section 2. The rate of an encoder is piecewise-constant between successive pairs of MPCRs.

We generate the rates as follows. We choose the rate such that the sequence of rate values is long-range dependent. Specifically, we use a fast-fourier transform method [14] to generate the fractional gaussian noise (FGN) sequence (an independent sequence for each source). We ignore values above the maximum rate to 15 Mbps and below the minimum rate (0 Mbps). The reason for this choice is discussed in the following section. We choose different values of mean and standard deviation for the generation procedure. When we generate an inter-MPCR interval $T_i$ and a corresponding rate $R_i$, the video source sends cells at a rate $R_i$ uniformly spaced in the interval $T_i$. Due to the ignoring of some rate values, the actual mean of the generated stream may be slightly greater or lesser than the input means. We later measure the actual mean rate and use it to calculate the efficiency metric.

Though each video source sends piecewise-CBR cell streams, the aggregate VBR rate need not be piecewise-CBR. It has a mean (SCR) which is the sum of all the individual means. Similarly, it has a maximum rate (PCR) which is close to the sum of the peak rates (15 Mbps) of the individual video streams. These quantities depend upon the number of video sources. In our model, we use N equal to 9 to ensure that the PCR is about 80% of total capacity. VBR is given priority at any link, i.e, if there is a VBR cell, it is scheduled for output on the link before any waiting ABR cells are scheduled. Further, since each video stream is long-range dependent, the composite VBR stream is also long-range dependent. Therefore, the composite VBR stream and the ABR capacity has high variance.

## 4.1 Observations on the Long-Range Dependent Traffic Generation Technique

The long-range dependent generation technique described in [14] can result in negative values and values greater than the maximum possible rate value. This occurs especially when the variance of the distribution is high (of the order of the mean itself). Fortunately, there are a few approaches in avoiding negative values and bounding values within a maximum in such sequences. We considered these approaches carefully before making a choice.

The first approach is to generate a long-range dependent sequence $x_1, x_2, ..., x_n$ and then use the sequence $e^{x_1}, e^{x_2}, ..., e^{x_n}$ in our simulation. The values $e^{x_i}$ is rounded off to the nearest integer. This method always gives zero or positive numbers. The new distribution still exhibits long-range dependence, though it is no longer a fractional gaussian noise (FGN) (like the originally generated sequence) [14]. Another problem is that all significant negative values are truncated to zero leading to an impulse at zero in the new probability density function (pdf). Further, the mean of the new sequence is not the exponentiated value of the old mean. This makes it difficult to obtain a sequence having a desired mean.

A second technique is to avoid exponentiation, but simply truncate negative numbers to zero. This approach again has the problem of the pdf impulse at zero. Also the mean of the entire distribution has increased.

The third technique is a variation of the second, which truncates the negative numbers to zero, but subtracts a negative value from the subsequent positive value. This approach is aimed to keep the mean constant. But, it not only has the side-effect of inducing a pdf impulse at zero, but also changes the shape of the pdf, thus increasing the probability of small positive values.

The fourth and final technique is to simply ignore negative values and values greater than the maximum. This approach keeps the shape of the positive part of the pdf intact while not introducing a pdf impulse at zero. If the number of negative values is small, the mean and variance of the distribution would not have changed appreciably. Further, it can be shown that the new distribution is still long-range dependent.

We choose the fourth approach (of ignoring negative values and values greater than the maximum) in our simulations.

sectionThe “N Source + VBR” Configuration

The “N Source + VBR” configuration shown in Figure 5 has a single bottleneck link shared by the N ABR sources and a VBR VC carrying the multiplexed stream. Each ABR source is a large (infinite) file transfer application using TCP. All traffic is unidirectional. All links run at 149.76 Mbps. The links traversed by the connections are symmetric i.e., each link on the path has the same length for all the VCs. In our simulations, N is 15 and the link lengths are 1000 km in WAN simulations. In satellite simulations, the feedback delay may be 550 ms (corresponds to a bottleneck after the satellite link) or 10 ms (corresponds to a bottleneck before the satellite link). This is illustrated in Figures 6 and 7 (section 7.3).

The individual link lengths determine the round trip time (RTT) and the feedback delay. Feedback delay is the sum of the delay for feedback from the bottleneck switch to reach the source and the delay for the new load from the sources to reach the switch. It is at least twice the one-way propagation delay from the source to the bottleneck switch. The feedback delay determines how quickly the feedback is conveyed to the sources and how quickly the new load is sensed at the switch.

For the video sources, we choose means and standard deviations of video sources to have three sets of values (7.5 Mbps, 7 Mbps), (10 Mbps, 5 Mbps) and (5 Mbps, 5 Mbps). This choice ensures that the variance in all cases is high, but the mean varies and hence the total VBR load varies. The number of video sources (N) is 9 which means that the maximum VBR load is 80% of 149.76 Mbps link capacity. As discussed later the effective mean and variance (after bounding the generated value to within 0 and 15 Mbps) may be slightly different and it affects the efficiency measure.

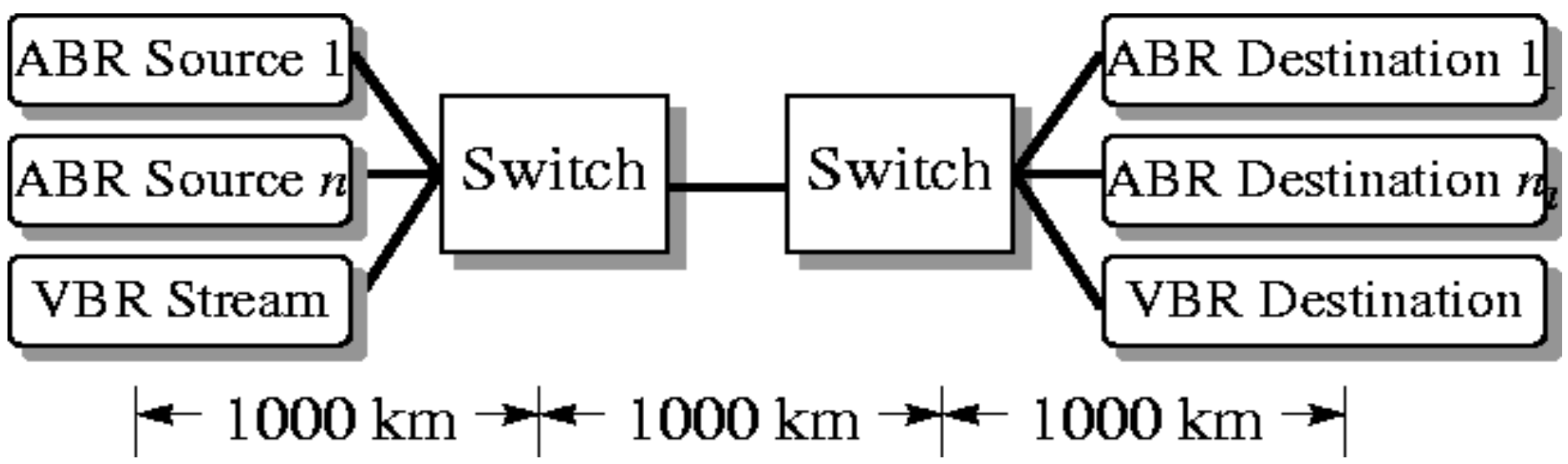

Figure 5: The "N Source + VBR" Configuration

The Hurst parameter which determines the degree of long-range dependence for each video stream is chosen as 0.8 [2].

We also compare results with prior results using an ON-OFF VBR model [15]. In this model, the ON time and OFF time are defined in terms of a "duty cycle" and a "period". A pulse with a duty cycle of d and period of p has an ON time of d×p and OFF time of (1-d)×p. When the duty cycle is 0.5, the ON time is equal to the OFF time. During the ON time, the VBR source operates at its maximum amplitude. The maximum amplitude of the VBR source is 124.41 Mbps (80% of link rate).

# 5 TCP and ERICA+ Parameters

We use a TCP maximum segment size (MSS) of 512 bytes. The window scaling option is used to obtain larger window sizes for our simulations. For WAN simulations we used a window of 16×64 kB or 1024 kB which is greater than the product of the round trip time (RTT) and the bandwidth yielding a result of 454,875 bytes at 121.3 Mbps TCP payload rate (defined below) when the RTT is 30 ms. For satellite simulations, we used a window size of $256 \times 34000 = 8.704 \times 10^6$ bytes which is sufficient for an RTT of 550 ms at 121.3 Mbps TCP payload rate.

TCP data is encapsulated over ATM as follows. First, a set of headers and trailers are added to every TCP segment. We have 20 bytes of TCP header, 20 bytes of IP header, 8 bytes for the RFC1577 LLC/SNAP encapsulation, and 8 bytes of AAL5 information, a total of 56 bytes. Hence, every MSS of 512 bytes becomes 568 bytes of payload for transmission over ATM. This payload with padding requires 12 ATM cells of 48 data bytes each. The maximum throughput of TCP over raw ATM is (512 bytes/(12 cells × 53 bytes/cell)) = 80.5%. Further in ABR, we send FRM cells once every Nrm (32) cells. Hence, the maximum throughput is $31/32 \times 0.805 = 78\%$ of ABR capacity. For example, when the ABR capacity is 149.76 Mbps, the maximum TCP payload rate is 116.3 Mbps. Similarly, for a MSS of 9140 bytes, the maximum throughput is 87% of ABR capacity.

We use a metric called "efficiency" which is defined as the ratio of the TCP throughput achieved to the maximum throughput possible. As defined above the maximum throughput possible is 0.78×(mean

ABR capacity). The efficiency is calculated as follows. We first measure the aggregate mean VBR rate (since it is not the sum of the individual mean rates due to bounding the values to 0 and 15 Mbps). Subtract it from 149.76 Mbps to get the mean ABR capacity. Then multiply the ABR capacity by 0.78 (or 0.87) to get the maximum possible throughput. We then take the ratio of the measured TCP throughput and this calculated value to give the efficiency.

In our simulations, we have not used the "fast retransmit and recovery" algorithms. Since there is no loss, these algorithms are not exercised.

The ERICA+ algorithm [4] uses five parameters. The algorithm measures the load and number of active sources over successive averaging intervals and tries to achieve 100% utilization with queueing delay equal to a target value. The averaging intervals end either after the specified length or after a specified number of cells have been received, whichever happens first. In our simulations, these values default to 500 ABR input cells or 5 ms. The other parameters are used to define a function which scales the ABR capacity in order to achieve the desired goals. These include a target queueing delay (T0, set to 500 microseconds), two curve parameters (a = 1.15 and b = 1.05), and a factor which limits the amount of ABR capacity allocated to drain the queues (QDLF = 0.5).

# 6 Simulation Results

## 6.1 Effect of High Variance and Total VBR Load

In this section, we present simulation results where we vary the mean and the standard deviation of the individual video sources such that the total variance is always high, and the total maximum VBR load varies.

In Table 1, and Table 2, we show the maximum queue length, the total TCP throughput, VBR throughput, ABR throughput, and efficiency for three combinations of the mean and standard deviation. Table 1 is for TCP MSS = 512 bytes, while Table 2 is for TCP MSS = 9140 bytes.

Table 1: Effect of Variance and VBR Load (MSS = 512, 10 sec simulation)

| Video Sources | | | ABR Metrics | | |
|---|---|---|---|---|---|
| # | Mean per-source rate (Mbps) | Standard Deviation (Mbps) | Max Switch Q (cells) | Total TCP Throughput | Efficiency ( % of Max throughput) |
| 1. | 5 | 5 | 6775 (1.8×F/b Delay) | 68.72 Mbps | 94.4% |
| 2. | 7.5 | 7 | 7078 (1.9×F/b Delay) | 59.62 Mbps | 94.1% |
| 3. | 10 | 5 | 5526 (1.5×F/b Delay) | 82.88 Mbps | 88.4% |

Observe that the measured mean VBR thoughput (column 6) is the same in corresponding rows of both the tables. This is because irrespective of ABR load, VBR load is given priority and cleared out first.

Table 2: Effect of Variance and VBR Load (MSS = 9140, 10 sec simulation)

| | Video Sources | | ABR Metrics | | |
|---|---|---|---|---|---|
| # | Mean per-source rate (Mbps) | Standard Deviation (Mbps) | Max Switch Q (cells) | Total TCP Throughput | Efficiency ( % of Max throughput) |
| 1. | 5 | 5 | 5572 (1.5×F/b Delay) | 77.62 Mbps | 95.6% |
| 2. | 7.5 | 7 | 5512 (1.5×F/b Delay) | 67.14 Mbps | 95.0% |
| 3. | 10 | 5 | 5545 (1.5×F/b Delay) | 56.15 Mbps | 95.6% |

Further, by bounding the MPEG-2 SPTS source rate values between 0 and 15 Mbps, we ensure that the total VBR load is about 80% of the link capacity.

For row 1, measured VBR throughput (column 6) was 56.44 Mbps (against $9 \times 5 = 45$ Mbps expected without bounding). For row 2, it was 68.51 Mbps (against 9 * 7.5 = 67.5 Mbps expected without bounding). For row 3, it was 82.28 Mbps(against 9 * 10 = 90 Mbps expected without bounding). Observe that when the input mean is higher, the expected aggregate value is lower and vice-versa.

The efficiency values (as defined in section 6) are calculated using these values of total VBR capacity. For example, in row 1 of Table 1, the ABR throughput is is 149.76 - 56.44 = 93.32 Mbps. For a MSS of 512, the maximum TCP thoughput is 78% of ABR throughput = 72.78 Mbps (not shown in the table). Given that TCP thoughput achieved is 68.72 Mbps (Column 5), the efficiency is 68.72/72.78 = 94.4%. For Table 2, since the MSS is 9140 bytes, the maximum TCP thoughput is 87% of ABR throughput as discussed in section 6, and this is the value used to compare the total TCP throughput against.

Observe that the efficiency achieved in all cases is high (above 90%) in spite of the high variance in ABR capacity. Also observe that the total TCP throughput is higher (as well as the efficiency) for TCP MSS = 9140 bytes in all cases.

The maximum queue length is controlled to about three times the feedback delay (or one round trip time) worth of queue. The feedback delay for this configuration is 10 ms, which corresponds to (10 ms) × (367 cells/ms) = 3670 cells worth of queue when the network is on the average overloaded by a factor of 2 (as is the case with TCP). The round-trip time for this configuration is 30 ms.

The queue length is higher when the mean per-source rate is lower (i.e., when the average ABR rate is higher). This is explained as follows. Whenever there is variance in capacity, the switch algorithm may make errors in estimating the average capacity and may overallocate rates temporarily. When the average ABR capacity is higher, each error in allocating rates will result in a larger backlog of cells to be cleared than for the corresponding case when the average ABR capacity is low. The combination of these backlogs may result in a larger maximum queue before the long-term queue reduction mechanism of the switch algorithm reduces the queues.

## 6.2 Comparison with ON-OFF VBR Results

In our earlier paper [15] and references therein, we had studied the behavior of TCP over ABR in the presence of ON-OFF VBR sources. We had studied ranges of ON-OFF periods from 1 ms through 100 ms. Further, we had looked at results where the ON period was not equal to the OFF period. The worst cases were seen in the latter simulations. However, with modifications to ERICA+ and a larger averaging interval we found that the maximum switch queue length was 5637 cells. This experiment has a duty cycle of 0.7 and a period of 20ms i.e., the ON time was 14 ms and the off time was 6 ms. Since we use the same switch algorithm parameters in this study, we can perform a comparison of the two studies.

We observe that, even after the introduction of the long-range dependent VBR model, the queues do not increase substantially (beyond one round trip worth of queues) and the efficiency remains high (around 90%). This is because the ERICA+ switch algorithm has been refined and tuned to handle variance in the ABR capacity and ABR demand. These refinements allow the convergence of the ABR queues, without compromising on the efficiency.

## 6.3 Satellite simulations with Short Feedback Delay

In this section and the next, we repeat the experiments with some links being satellite links. In the first set of simulations, we replace the bottleneck link shared by 15 sources with a satellite link as shown in Figure 6. The links from the second switch to the destination nodes are 1 km each. The total round trip time is 550 ms, but the feedback delay remains 10 ms.

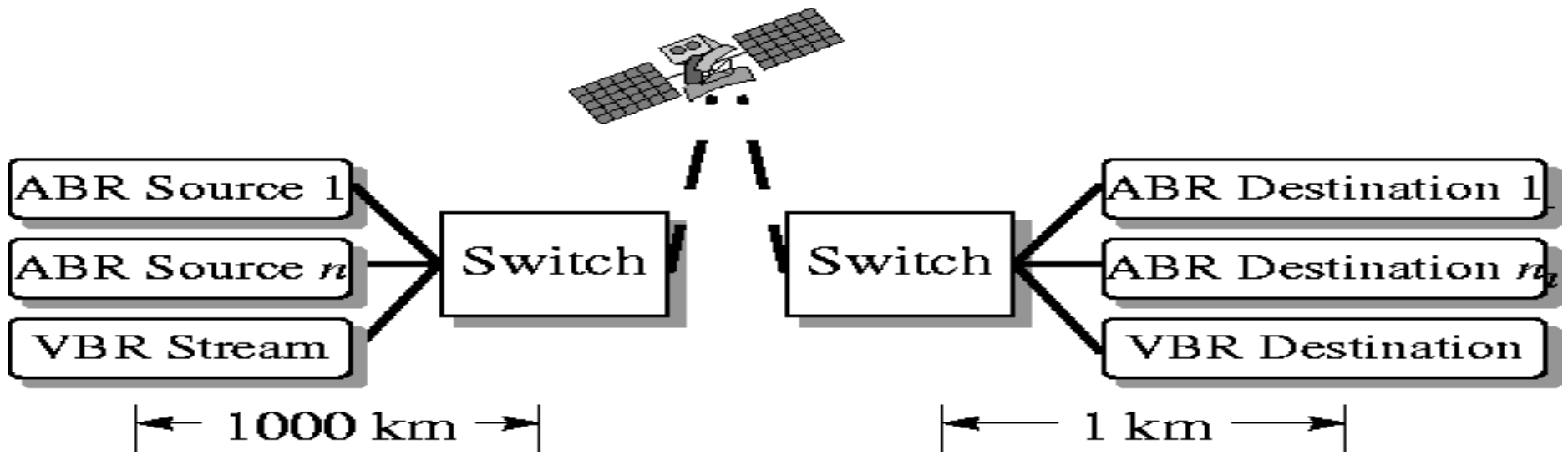

Figure 6: The "N Source + VBR" Configuration with a satellite link

Table 3 and Table 4 (similar to Tables 1 and 2) show the maximum switch queue length, the total TCP throughput, VBR throughput, ABR throughput, and efficiency for three combinations of the mean and standard deviation. Table 3 is for TCP MSS = 512 bytes, while Table 4 is for TCP MSS = 9140 bytes.

Note that the TCP startup time in this configuration is large because the round trip time (550 ms) is large and TCP requires multiple round trips to be able to use its full capacity. However, the effect on total TCP throughput is minimal since there is no loss and the feedback delays are small (10 ms) compared to round trip time, allowing ABR to control sources more effectively. Throughputs are high, and efficiency values are high.

The tables shows that maximum queues are small (in the order of three times the feedback delay), irrespective of the mean and variance. In such satellite configurations, we observe that the feedback delay is the dominant factor (over round trip time) in determining the maximum queue length. As discussed earlier, one feedback delay of 10 ms corresponds to 3670 cells of queue for TCP.

Table 3: Max Queues for Satellite Networks with Short Feedback Delay (MSS=512, 170 sec)

| | Video Sources | | ABR Metrics | | |
|---|---|---|---|---|---|
| # | Mean per-source rate (Mbps) | Standard Deviation (Mbps) | Max Switch Q (cells) | Total TCP Throughput | Efficiency ( % of Max throughput) |
| 1. | 5 | 5 | 5545 (1.5×f/b delay) | 68.09 | 92.9% |
| 2. | 7.5 | 7 | 4416 (1.2×f/b delay) | 59.16 | 82.5% |
| 3. | 10 | 5 | 4064 (1.1×f/b delay) | 47.39 | 86.7% |

Table 4: Max. Queues for Satellite Networks with Short Feedback Delay (MSS=9140, 170 sec)

| | Video Sources | | ABR Metrics | | |
|---|---|---|---|---|---|
| # | Mean per-source rate (Mbps) | Standard Deviation (Mbps) | Max Switch Q (cells) | Total TCP Throughput | Efficiency ( % of Max throughput) |
| 1. | 5 | 5 | 5759 (1.6×f/b delay) | 72.18 | 88.3% |
| 2. | 7.5 | 7 | 11366 (3.1×f/b delay) | 67.23 | 84.1% |
| 3. | 10 | 5 | 13105 (3.6×f/b delay) | 57.69 | 94.6% |

## 6.4 Satellite simulations with Long Feedback Delay

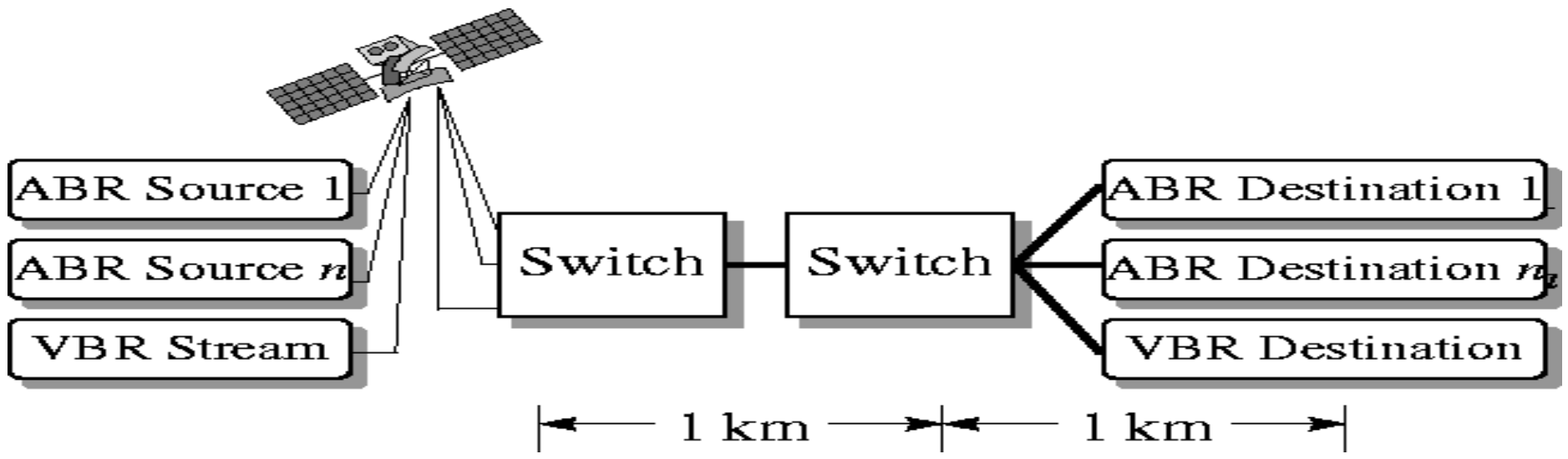

Figure 7: The "N Source + VBR" Configuration with satellite links and long feedback delays

In our second set of satellite simulations, we examine the effect of longer feedback delays. Consider a switch A at the end of a satellite link or a switch downstream of A. It will have a feedback delay of about 550 ms. This is the scenario we model. We form a new configuration as shown in Figure 7 by replacing the links in the feedback path to sources with satellite link. All other links are of length 1 km each. As a result, the round trip time and the feedback delay are both approximately equal to 550 ms.

Tables 5 and 6 (similar to Tables 1 and 2) show the maximum switch queue length, the total TCP throughput, VBR throughput, ABR throughput, and efficiency for three combinations of the mean and standard deviation. Table 5 is for TCP MSS = 512 bytes, while Table 6 is for TCP MSS = 9140 bytes.

Table 5: Max Queues for Satellite Networks with Long Feedback Delay (MSS=512, 170 sec)

| | Video Sources | | ABR Metrics | | |
|---|---|---|---|---|---|
| # | Mean per-source rate (Mbps) | Standard Deviation (Mbps) | Max Switch Q (cells) | Total TCP Throughput | Efficiency ( % of Max throughput) |
| 1. | 5 | 5 | 299680 (1.5×f/b delay) | 58.25 | 79.60% |
| 2. | 7.5 | 7 | 186931 (0.9×f/b delay) | 49.50 | 80.53% |
| 3. | 10 | 5 | 156232 (0.8×f/b delay) | 44.16 | 80.93% |

Table 6: Max Queues for Satellite Networks with Long Feedback Delay (MSS=9140, 170 sec)

| | Video Sources | | ABR Metrics | | |
|---|---|---|---|---|---|
| # | Mean per-source rate (Mbps) | Standard Deviation (Mbps) | Max Switch Q (cells) | Total TCP Throughput | Efficiency ( % of Max throughput) |
| 1. | 5 | 5 | 215786 (1.1×f/b delay) | 65.96 | 80.81% |
| 2. | 7.5 | 7 | 233423 (1.2×f/b delay) | 58.76 | 85.71% |
| 3. | 10 | 5 | 144362 (0.7×f/b delay) | 49.67 | 81.61% |

Observe that the queue lengths are quite large, while the total TCP throughput and efficiency are smaller (by 6-13%) compared to the values in Tables 1 and 2 (1000 km feedback delay cases) respectively. The total queue is still a small multiple of the feedback delay or RTT (a feedback delay of 550 ms corresponds to 201850 cells). This indicates that satellite switches need to provide at least so much buffering to avoid loss on these high delay paths. A point to consider is that these large queues should not be seen in downstream workgroup or WAN switches, because they will not provide so much buffering. Satellite switches can isolate downstream switches from such large queues by implementing the VSVD option as described in our previous contribution [17].

# 7 Summary

Compressed video sources exhibit long-range dependence in the traffic patterns they generate. In this paper, we briefly survey VBR video modeling techniques, the MPEG-2 over ATM approach, and propose a model for MPEG-2 video over VBR which incorporates the long-range dependence property in compressed video. We have shown how to combine the fractional guassian noise sequences generated using fast fourier transforms can be used to produce long-range dependent traffic which models multiplex MPEG-2 video sources over VBR. The effect of this long-range dependent traffic over VBR is to introduce high variance in the ABR capacity. However, a responsive switch scheme like ERICA+ is sufficient to handle this variance in ABR capacity. This results in controlled ABR queues and high utilization. The maximum ABR queue length is a function of the feedback delay and round trip time. This implies that switches terminating satellite links should provide buffers proportional to the length of the satellite link in order to deliver high performance. Further, if they implement the VSVD option, they can isolate downstream workgroup switches from the effects of the long delay satellite path.

[4] Throughout this section, AF-TM refers to ATM Forum Traffic Management sub-working group contributions.

[5] All our papers and ATM Forum contributions are available through http://www.cis.ohio-state.edu/~jain/